\documentclass{ws-ijmpd}

\begin{document}

\markboth{Saibal Ray, Farook Rahaman and Utpal Mukhopadhyay}
{Scenarios of cosmic string with variable cosmological constant}

%
\catchline{}{}{}{}{}
%

\title{Scenarios of cosmic string with variable cosmological constant}

\author{\footnotesize Saibal Ray}
\address{Department of Physics, Barasat Government College, Kolkata 700 124,
North 24 Parganas, West Bengal, India \& IUCAA, Post Bag 4, Pune
411 007, India; saibal@iucaa.ernet.in}

\author{Farook Rahaman}
\address{Department of Mathematics, Jadavpur University,
  Kolkata 700 032, West Bengal, India}

\author{Utpal Mukhopadhyay}
\address{Satyabharati Vidyapith, Barasat, North 24 Parganas,
Kolkata 700 126, West Bengal, India}

\maketitle

\begin{history}
\received{Day Month Year} \revised{Day Month Year}
\end{history}

\begin{abstract}
Exact solutions of the Einstein field equations with cosmic string
and space varying cosmological constant, viz., $\Lambda=
\Lambda(r)$, in the energy-momentum tensors are presented. Three
cases have been studied: where variable cosmological constant (1)
has power law dependence, (2) is proportional to the string fluid
density, and (3) is purely a constant. Some cases of interesting
physical consequences have been found out such that (i) variable
cosmological constant can be represented by a power law of the
type $\Lambda=3r^{-2}$, (ii) variable cosmological constant and
cosmic string density are interdependent to each other according
to the relation $\Lambda=-8\pi{\rho}_s$, and (iii) cosmic string
density can be scaled by a power law of the type
${\rho}_s=r^{-2}$. It is also shown that several known solutions
can be recovered from the general form of the solutions obtained
here.

\end{abstract}

\section{Introduction}
 The cosmological constant $\Lambda$ has a fairly long history of
acceptance and rejection due to its peculiar properties and
behaviour. Historically, it was first proposed by
Einstein~\cite{ein1917} as repulsive pressure to achieve a
stationary universe for his theory of general relativity. But,
later on, theoretical work of Friedmann~\cite{fri1922} without
cosmological constant and observational result of
Hubble~\cite{hub1929} with galactic law of red-shifts - all did
indicate towards an expanding universe paradigm. This situation
forced Einstein to abandon the concept of introducing $\Lambda$ in
his gravitational field equations. Thus what was rejected by
Einstein the same was embraced by Friedmann and Hubble. This is
the first phase for its weird history of ups and downs in the
first quarter of nineteenth century.

However, the debate went on with its non-existence and even
existence. Actually Zel'Dovich~\cite{zel1968} in his most
innovative way revived the issue of cosmological constant
$\Lambda$ by identifying it with the vacuum energy density due to
quantum fluctuations. Therefore, people started thinking about
$\Lambda$ once again with a new outlook. If it does exist then
what are the features and implications of it on the physical
systems are available in these review
articles~\cite{wei1989,car1992,sah2000,pad2003}. Following all
these what we are really interested to mention here that survival
of cosmological constant $\Lambda$ was occurring slowly and
getting gradually a strong theoretical footing. However, the
resurrection came in a bold way through the observational evidence
of high redshift Type Ia supernovae~\cite{per1998,rie1998} for a
small decreasing value of cosmological constant
($\Lambda_{present}\leq 10^{-56} cm^{-2}$) at the present epoch.
It is now commonly believed  by the scientific community that via
cosmological constant a kind of repulsive pressure, {\it dark
energy}, is responsible for this present state of accelerating
Universe. This is because of the fact that the $\Lambda$-CDM model
agrees closely with all most all the established cosmological
observations, the latest being the 2005 Supernova Legacy Survey
(SNLS). According to the first year data set of SNLS {\it dark
energy} behaves like the cosmological constant to a precision of
$10$ per cent~\cite{ast2006}.

Obviously, this small decreasing value of cosmological constant
indicates that instead of a strict constant then $\Lambda$ could
be a function of space and time coordinates or even of both. It
is, therefore, argued that the solutions of Einstein's field
equations with variable $\Lambda$ will have a wider range and the
roll for scalar $\Lambda$ viz., $\Lambda=
\Lambda(r)$, in astrophysical problems in
relation to the nature of local massive objects like galaxies,
will be as much significant as in
cosmology~\cite{nar1991,tiw2000,ray2004}. In this line of
approach, the field equations including variable $\Lambda$ have been solved
in various cosmological situations (i.e. $\Lambda=
\Lambda(t)$) as well as astrophysical situations (i.e. $\Lambda=
\Lambda(r)$) by several
workers~\cite{sak1968,gun1975,lau1985,ber1986a,ber1986b,reu1987,pee1988,ber1989,che1990,sis1991,kal1992,bee1993,tiw1996,bro2003,wan2005,sol2006}.

Now, string theory being a major tool for understanding nature
involving physical systems a string theoretic approach towards
astrophysical and cosmological realm thought to be useful from the
very beginning of the advent of the theory.
Letelier~\cite{let1979,let1981} studied the spherical symmetric
gravitational field produced by a bunch of strings by replacing
the motion of dust clouds and perfect fluids in terms of string
clouds and string fluids. Later on Vilenkin~\cite{vil1984}
proposed a model of a universe composed of string clouds. Einstein
field equations have also been solved by several workers from
string theoretic point of view for a single cosmic
string~\cite{got1985,his1985}, two moving straight
strings~\cite{got1991} and $N$ straight strings moving on a
circle~\cite{kab1992}. Dabrowski and Stelmach~\cite{dab1989}
considered Friedman universe filled with mutually noninteracting
pressureless dust, radiation, cosmological constant and dense
system of strings. There are also several other very recent models
where the concepts of string have been considered, e.g., rolling
tachyon from the string theory~\cite{kre2004}, a string-inspired
scenario associated with a rolling massive scalar field on
D-branes~\cite{gar2004}, exotic matter, either in the form of
cosmic strings or struts with negative energy density also has
been considered in connection to the matter content of an
additional thin rotating galactic disk~\cite{vog2005} and a fluid
of strings along with the cosmological constant, as a candidate of
dark energy, has considered by Capozziello et al.~\cite{cap2006}
to get a viable scenario of the present status of the Universe.

So, it is not unnatural to investigate the effect of space-variable $\Lambda$ on
an anisotropic static spherically symmetric source composed of
string fluid. Therefore, the present investigation is the
generalization of the work of Tiwari and Ray~\cite{tiw1996} with
string fluid replacing perfect fluid. The main purpose of the
investigations are: firstly, to search for some kind of link
between the approaches with matter and string theoretic one;
secondly, interrelationship, if exist, between the string and
cosmological constant; and thirdly, to find out the extra features
which appear due to the inclusion of varying cosmological constant
in string fluid.

The paper is organized as follows: in Section 2 the general
relativistic Einstein field equations are presented where the
cosmological constant and string fluid are included in the
energy-momentum tensors. Section 3 deals with their solutions in
various cases and subcases under this modified theory of general
relativity. We include two Sections 4 and 5 regarding solar system
and effective attractive correction to the Newtonian force due to
cosmic string density admitting varying $\Lambda$. Some salient
features of the present model are discussed in the concluding
Section~6.

\section{Einstein field equations}
The Einstein field equations are given by
\begin{equation}
{G^{i}}_{j} = {R^{i}}_{j} - \frac{1}{2}{{g^{i}}_{j}} R =
-8\pi{T^i}_j - {g^i}_j \Lambda, \quad (i,j = 0, 1, 2, 3)
\end{equation}
where in relativistic units both $G$ and $c$ of $\kappa(=8\pi
G/c^4)$ are equal to $1$. Here, we shall assume the cosmological
constant $\Lambda$ to be space-varying, i.e.,
$\Lambda=\Lambda(r)$. It is to be noted here that the left hand
side of the Einstein field equations express the geometrical
structure of the space-time whereas the right hand side is the
representative of the matter content. We have, therefore,
purposely put the non-zero $\Lambda$ term in the right to make its
status as physical one such that the conservation law in the
present case takes the form as follows~\cite{pee1988}
\begin{equation}
{8\pi{T^i}_j};i =- \Lambda;j.
\end{equation}

Let us consider a spherically symmetric line element
\begin{equation}
ds^2 = g_{ij}dx^{i}dx^{j} = e^{\nu(r)} dt^2 - e^{\lambda(r)} dr^2
- r^2 (d \theta ^2 + sin^2 \theta d\phi^2)
\end{equation}
where $\nu$ and $\lambda$ are the metric potentials and are
function of the space coordinate $r$ only, such that $\nu=\nu(r)$
and $\lambda=\lambda(r)$ .

Now, if we assume that (1) the fluid source is a perfect fluid of
finite-length straight strings and (2) the individual strings do
orient themselves in a perfectly radial direction, then the
energy-momentum tensors reduce to
\begin{equation}
{T^t}_t ={T^r}_r \quad and \quad {T^{\omega}}_{\omega} =q
\end{equation}
where $\omega$ stands for both the angular coordinates $\theta$
and $\phi$ related to the metric (3). In this context it is to be
noted here that the above assumptions are quite relevant to the
results of Gott~\cite{got1985} and Hiscock~\cite{his1985} that a
straight ideal string has vanishing gravitational mass due to the
gravitational effect of tension which exactly cancels the effect
of mass. Therefore, strings do not produce any gravitational force
and can be thought of exactly straight and radially
oriented~\cite{let1979,sol1995}.

Now, to solve the Einstein's gravitational field equations
uniquely one has to specify some relationship between the
energy-momentum tensors. Let us, therefore, assume that the
transverse pressure part of the string energy-momentum tensor
${T^t}_t$ is proportional to the angular part of the
energy-momentum tensors ${T^{\omega}}_{\omega}$. Then, from the
above equation (4), we get
\begin{equation}
{T^t}_t ={T^r}_r=-\alpha {T^{\omega}}_{\omega}
\end{equation}
where $\alpha$ is a dimensionless constant of proportionality.
This assumption (5), as mentioned in the previous discussion,
indicates the perfect radial orientation of the strings. Here,
however, we would like to mention that this algebraic form of
energy-momentum tensors was earlier considered by
Gliner~\cite{gli1966} and Petrov~\cite{pet1969} in mathematical
context and later on by Dymnikova~\cite{dym1992} and
Soleng~\cite{sol1994} associated with anisotropic vacuum
polarization in a spherically symmetric space-time.

By virtue of the equations (3) and (5), the field equations (1)
can explicitly be expressed as
\begin{equation}
e^{-\lambda} \left( \frac{\lambda^{\prime}}{r} - \frac{1}{r^{2}}
\right) + \frac{1}{r^{2}} = 8\pi {\rho}_s + \Lambda,
\end{equation}
\begin{equation}
e^{-\lambda} \left( \frac{\nu^{\prime}}{r} + \frac{1}{r^{2}}
\right) - \frac{1}{r^{2}} = -8\pi {\rho}_s - \Lambda,
\end{equation}
\begin{equation}
e^{-\lambda} \left [\frac{\nu^{{\prime}{\prime}}}{2} +
\frac{{\nu^{\prime}}^{2}}{4} - \frac{{\nu^{\prime}
\lambda^{\prime}}}{4} + \frac{(\nu^{\prime} - \lambda^{\prime} )}{
2r}\right]= \frac{8\pi}{\alpha} {\rho}_s - \Lambda.
\end{equation}
The above field equations, at a glance, give some clue in between
the metric potentials. Therefore, addition of the equations (6)
and (7) immediately provides us the following relation
\begin{equation}
\nu=- \lambda
\end{equation}
which is equivalent to $g_{00}g_{11}=-1$, between the metric
potentials of the metric equation (3). A coordinate-independent
statement of this relation have been obtained by Tiwari, Rao and
Kanakamedala~\cite{tiw1984} by using the eigen values of the
Einstein tensor ${G^{i}}_{j}$ expressed in equation (1). It is
also interesting to note that $g_{00}g_{11} = -1$ can be expressed
in terms of the energy-momentum tensors ${T^{1}}_{1} =
{T^{0}}_{0}$.

The energy conservation law, in the present situation, is given by
\begin{equation}
\frac{d}{dr}\left[-{\rho}_s - \frac{\Lambda}{8\pi}\right] =
\frac{2}{r}\left[{\rho}_s +
\frac{{\rho}_s}{\alpha}\right]=\frac{\beta}{r}{\rho}_s
\end{equation}
where
\begin{equation}
\beta=2\left(1 + \frac{1}{\alpha}\right).
\end{equation}
Then, from the equation (10), we get
\begin{equation}
\frac{d{\rho}_s}{dr}+ \frac{\beta}{r}{\rho}_s=-
\frac{1}{8\pi}\frac{d\Lambda}{dr}
\end{equation}
which, on multiplication by $r^{\beta}$ and then integrating it,
reduces to
\begin{equation}
{\rho}_s r^{\beta}=\frac{A}{8\pi}-
\frac{1}{8\pi}\int{\Lambda}^{\prime}r^{\beta}dr
\end{equation}
where $A/8\pi$ is an integration constant.

\section{The solutions}
To obtain some explicit simplified results, let us investigate the
field equations under the following special assumptions.

\subsection{The case for ${\Lambda}^{\prime}\propto r^{-\beta}$}

Let us assume that
\begin{equation}
{\Lambda}^{\prime}=B r^{-\beta}
\end{equation}
where $B$ is a proportional constant.

By the use of this assumption (14), we get from the equation (13)
\begin{equation}
{\rho}_s r^{\beta}=\frac{A}{8\pi}- \frac{1}{8\pi}\int Bdr,
\end{equation}
so that the energy density can easily be obtained as
\begin{equation}
{\rho}_s =\frac{A}{8\pi}r^{-\beta}- \frac{B}{8\pi}r^{1-\beta}.
\end{equation}
Again, on integration, equation (14) yields
\begin{equation}
\Lambda = \frac{B}{1-\beta}r^{1-\beta}.
\end{equation}
Using the equations (16) and (17) we get from the equation (6)
\begin{equation}
e^{-\lambda} \left( \frac{\lambda^{\prime}}{r} - \frac{1}{r^{2}}
\right) + \frac{1}{r^{2}} =Ar^{-\beta}+
\frac{B}{1-\beta}r^{1-\beta}- Br^{1-\beta},
\end{equation}
so that after simplification it takes the following differential
form
\begin{equation}
d(re^{-\lambda})=dr - Ar^{2-\beta}dr + Br^{3-\beta}
\left[1-\frac{1}{1-\beta}\right]dr.
\end{equation}
The above equation (19) on integration provides the form for the
metric potentials as
\begin{equation}
e^{-\lambda}=1-\frac{2M}{r} - \frac{A}{3-\beta}r^{2-\beta}-
\frac{B\beta}{(1-\beta)(4-\beta)}r^{3-\beta} = e^{\nu}
\end{equation}
where $2M$ is an integration constant.

Thus the general form of the space-time can be given by
\begin{eqnarray}
ds^2 = \left[1-\frac{2M}{r} - \frac{A}{3-\beta}r^{2-\beta}-
\frac{B\beta}{(1-\beta)(4-\beta)}r^{3-\beta}\right] dt^2
\nonumber\\- \left[1-\frac{2M}{r} - \frac{A}{3-\beta}r^{2-\beta}-
\frac{B\beta}{(1-\beta)(4-\beta)}r^{3-\beta}\right]^{-1} dr^2
\nonumber\\- r^2 (d \theta ^2 + sin^2 \theta d\phi^2).
\end{eqnarray}

\subsubsection{$\alpha=-2$}

For this subcase when $\beta=1$ readily makes $\Lambda$, via
equation (17), to be infinite and hence is not a well-defined
quantity. However, for $\beta=-1$ one can get
\begin{equation}
\Lambda=\frac{B}{2}r^2.
\end{equation}
For a particular choice of the values for $\beta=3$ and $B=-6$,
therefore, equation (17) gives
\begin{equation}
\Lambda=\frac{3}{r^2}.
\end{equation}
This result is similar to the one as obtained by Krisch and
Glass~\cite{kri2003} when $r$ is recognized as the cosmic scale
factor in connection to a toroidal fluid solution embedded in a
locally anti-de Sitter exterior.

\subsubsection{$\alpha=-1$}

For this, we get from the equation (11) the value for $\beta$ as
zero, which reduces the equation (10) in the form
\begin{equation}
\frac{d}{dr}\left[{\rho}_s + \frac{\Lambda}{8\pi}\right] = 0.
\end{equation}
This immediately yields
\begin{equation}
{\rho}_s + \frac{\Lambda}{8\pi}=C
\end{equation}
where $C$ is an integration constant. Hence, from the equation
(6), after integrating it and using results of (9) and (25), we
get
\begin{equation}
e^{-\lambda}=1-\frac{2M}{r}- \frac{8\pi C}{3}r^2 = e^{\nu}.
\end{equation}
From the above equation (26) we find that, for the appropriate
choice of value of the constant as $C=0$, it turns out to be
Schwarzschild solution. A comparison of the equations (20) and
(26), via equation (25), immediately reveals that $A=8\pi
C=\Lambda + 8\pi{\rho}_s$ for $\beta=0$. Therefore, choosing
suitably the values of $A=C=0$ one can obtain
\begin{equation}
\Lambda = - 8\pi{\rho}_s.
\end{equation}
This means that, even in a restricted case, vacuum energy density
is dependent on cosmic string density and vice versa. It is
worthwhile to note that, in the context of the inflationary
cosmology~\cite{gut1981,sat1981,lin1982,alb1982,lin1983}, the
above relation (27) can be expressed as $p_s = - {\rho}_s$, where
$p_s$, now equivalent to $\Lambda/8\pi$, is the fluid pressure of
string. The equation of state of this type $p_s=\gamma{\rho}_s$
with $\gamma=-1$ implies that the matter distribution is in
tension and hence the matter is known, in the literature, as a
`false vacuum' or `degenerate vacuum' or
`$\rho$-vacuum'~\cite{dav1984,blo1984,hog1984,kai1984}.

Now two cases may be considered here -\\ {\it Case I:} When the
condition is ${\rho}_s>0$ and $\Lambda<0$, we get the positive
energy density of string with negative vacuum energy density. We
emphasize here that in the cosmological context $\Lambda$ positive
is related to the repulsive pressure. $\Lambda$ being negative, as
evident from the equation (27), this case does not correspond to
the present state of the accelerating Universe rather it is
related to a collapsing situation~\cite{car2003}.

{\it Case II:} When the condition is ${\rho}_s<0$ and $\Lambda>0$,
we get the negative energy density of string with positive vacuum
energy density. This case of positive $\Lambda$ and hence negative
pressure, therefore, indicates towards a state of acceleration.
This positive $\Lambda$ as appears in the form of the negative
pressure try to expand the space-time curvature in the outward
direction and thus does play the role for {\it dark energy} which
is responsible for the present status of the cosmic
acceleration~\cite{per1998,rie1998}. It is to be noted here that
the concept of negative energy density of string is not
unrealistic in the realm of string theory and have been
extensively studied by several
workers~\cite{sen2000,vol2002,hor2003,her2003,her2004}.

\subsection{The case for $\Lambda \propto {\rho}_s$}

Let us assume here that
\begin{equation}
\Lambda =8\pi D {\rho}_s
\end{equation}
where $D$ is a constant of proportionality.

From the above assumption (28), after differentiating it, we get
\begin{equation}
\frac{d\Lambda}{dr}=8\pi D\frac{d{\rho}_s}{dr}
\end{equation}
for which the equation (12) reduces to
\begin{equation}
\frac{d{\rho}_s}{dr}+
\frac{\beta}{r}{\rho}_s=-D\frac{d{\rho}_s}{dr}.
\end{equation}
On integration this yields
\begin{equation}
{\rho}_s=D_0 r^{-\beta/(1+D)}.
\end{equation}
Therefore, by the use of the equations (28) and (31), we get from
the equation (6)
\begin{equation}
e^{-\lambda}=1-\frac{2M}{r} - \frac{8\pi
D_0(1+D)^2}{3+3D-\beta}r^{(2+2D-\beta)/(1+D)} = e^{\nu}.
\end{equation}
Thus the general form of the space-time can be written as
\begin{eqnarray}
ds^2 = \left[1-\frac{2M}{r} - \frac{8\pi
D_0(1+D)^2}{3+3D-\beta}r^{(2+2D-\beta)/(1+D)}\right] dt^2
\nonumber\\- \left[1-\frac{2M}{r} - \frac{8\pi
D_0(1+D)^2}{3+3D-\beta}r^{(2+2D-\beta)/(1+D)}\right]^{-1} dr^2
\nonumber\\- r^2 (d \theta ^2 + sin^2 \theta d\phi^2).
\end{eqnarray}

\subsubsection{$D=-1$}

This subcase, obviously, goes back to the subcase $3.1.2$ for
$A=C=0$ so that the explanation in connection to the negative
string density and cosmic acceleration is also valid here. The
Schwarzschild vacuum solution also can be recovered here as a
special case in the form given by
\begin{eqnarray}
ds^2 = \left[1-\frac{2M}{r}\right] dt^2 -
\left[1-\frac{2M}{r}\right]^{-1} dr^2 \nonumber\\- r^2 (d \theta
^2 + sin^2 \theta d\phi^2).
\end{eqnarray}
This space-time is associated with a particle of mass $M$ centered
at the origin of the spherical system surrounded by a cloud of
strings of density ${\rho}_s=D_0 r^{-\infty}$, for $D_0 \neq 0$.
Obviously, there does not exist any string around the spherical
configuration except at the points where $r = \pm 1$. At $r=0$ the
string density and space-time both blow up and thus {\it
singularity} arises in the solution. According to
Soleng~\cite{sol1995}, however, this singularity at the origin is
not a unique feature for string fluid model the mass being
surrounded by a string fluid.

\subsubsection{$D=0$}

For this subcase the cosmological parameter $\Lambda$ vanishes by
virtue of assumption (28) and hence the equation (31) reduces to
\begin{equation}
{\rho}_s=D_0 r^{-\beta}.
\end{equation}
This result can easily be recognized as the one obtained by Turok
and Bhattacharjee~\cite{tur1984} and also by Kibble~\cite{kib1986}
where $2\leq \beta \leq 3$ in their case. If we now choose
$\alpha>>1$ then for this very large value of $\alpha$, when
$\beta=2$, the equation (31) becomes
\begin{equation}
{\rho}_s \propto r^{-2}.
\end{equation}
This suggests that string-dominated universe expands more rapidly
than matter- or radiation-dominated one where energy density
varies, respectively, as ${\rho}_m \propto r^{-3}$ and ${\rho}_r
\propto r^{-4}$~\cite{spe1997}. This case $\beta=2$ and the astronomical
constraint on string-dominated universe have been investigated by
Gott and Rees~\cite{got1987}. However, Dabrowski and
Stelmach~\cite{dab1989} prefer this case $\beta=2$ as it
corresponds to the set of randomly oriented straight strings or to
the tangled network of strings which conformally stretches by
expansion. According to them unlike $2\leq \beta \leq 3$ this case
is particularly interesting because it allows treatment of all
type of components of the universe simultaneously in an analytic
way and also gives some aspects of observational problems in the
universe with strings. Vilenkin~\cite{vil1985} and
Soleng~\cite{sol1995} also exploit this type of string cloud with
energy density ${\rho}_s \propto r^{-2}$.

In this connection we are interested to point out that the present
model with ${\rho}_s \propto r^{-\beta}$ is more general than
others as mentioned above and can be applicable, at least
theoretically, with out any constrains like $2\leq \beta \leq 3$.
Another point to note here is that this result appears in the form
of a solution of the Einstein field equations not as an {\it ad
hoc} assumption.

\subsection{The case for $\Lambda =constant $}

For this trivial case of erstwhile cosmological constant, the
equation (13) becomes
\begin{equation}
{\rho}_s=\frac{A}{8\pi} r^{-\beta}.
\end{equation}
This is again the same relation (35) for $A=8\pi D_0$. Therefore,
we can recover the relation ${\rho}_s \propto r^{-2}$ from here
also for $\alpha=\infty$, $\beta=2$ and hence the explanation of
this can be made in a similar way as before.

By the use of above expression in the equation (6) the metric
potential can now be given by
\begin{equation}
e^{-\lambda}=1-\frac{2M}{r} - \frac{\Lambda}{3}r^2
-\frac{A}{3-\beta}r^{2-\beta} = e^{\nu}.
\end{equation}
Thus the general form of the space-time can be written as
\begin{eqnarray}
ds^2 = \left[1-\frac{2M}{r} - \frac{\Lambda}{3}r^2
-\frac{A}{3-\beta}r^{2-\beta}\right] dt^2 \nonumber\\ -
\left[1-\frac{2M}{r} - \frac{\Lambda}{3}r^2
-\frac{A}{3-\beta}r^{2-\beta}\right]^{-1} dr^2 \nonumber\\ - r^2
(d \theta ^2 + sin^2 \theta d\phi^2).
\end{eqnarray}

\subsubsection{$\alpha=-1$}

For this subcase we get $\beta=0$ and hence the equation (38)
reduces to
\begin{equation}
e^{-\lambda}=1-\frac{2M}{r} - \frac{\Lambda}{3}r^2 -\frac{A}{3}r^2
= e^{\nu}.
\end{equation}
One can observe that for $A=0$ the above space-time becomes the
Schwarzschild-de Sitter vacuum solution which again, in the
absence of $\Lambda$ reduces to the Schwarzschild vacuum solution
as in the case of Soleng~\cite{sol1995} for $\alpha = -1$ and $0$
respectively. Moreover, in this case the string density becomes a
constant quantity as given by
\begin{equation}
{\rho}_s=\frac{A}{8\pi}
\end{equation}
which means that the space-time associated with the particle of
mass $M$ is surrounded by a spherical cloud of constant string
density here.

\subsubsection{$\alpha=1$}

In this subcase of $\beta=4$, we get the metric as
\begin{eqnarray}
ds^2 = \left[1-\frac{2M}{r}- \frac{\Lambda}{3}r^2 +
\frac{A}{r^2}\right] dt^2 \nonumber\\- \left[1-\frac{2M}{r}-
\frac{\Lambda}{3}r^2 + \frac{A}{r^2}\right]^{-1} dr^2 \nonumber\\-
r^2 (d \theta ^2 + sin^2 \theta d\phi^2).
\end{eqnarray}

If we now smoothly match this space-time with that of the
Reissner-Nordstr{\"o}m on the boundary of the spherical object of
radius $R$ such that $R=r$, then we get
\begin{equation}
A=Q^2
\end{equation}
in the absence of $\Lambda$. Therefore, the constant $A$ can
easily be identified with the square of the charge contained in
the sphere and hence the above solution provides a
Reissner-Nordstr{\"o}m black hole surrounded by strings of fluid
density ${\rho}_s=(A/8\pi) r^{-4}$. However, in the case of
$\Lambda \neq 0$ this turns into a Reissner-Nordstr{\"o}m-de
Sitter universe~\cite{sol1995}.

\section{Solar system with cosmic string density admitting
varying $\Lambda$}

To study the gravitational effects of cosmic string density
admitting varying $\Lambda$ on planetary motion, we discuss the
perihelion precession of planet. To find perihelion shift, we use
usual geodesic equation for a massive particle (e.g. planet),
which can be obtained in a standard manner for the spherically
symmetric metric of the form
\begin{equation}
                ds^2 = f(r)dt^2 - \frac{1}{f(r)}dr^2-r^2 d\theta^2-r^2 \sin ^2 \theta d\phi^2
\end{equation}

as~\cite{wei1972}

\begin{equation}
             \frac{1}{r^4}\left[\frac{dr}{d\phi}\right]^2 =
             \frac{E^2}{p^2}- \frac{f}{p^2}-  \frac{f}{r^2}.
\end{equation}

Here, $f\dot{t} = E$, $r^2\dot{\theta} = p$ where $E$ and $p$ are
the energy and momentum of the particle respectively and over dot
implies differentiation with respect to affine parameter $s$.

By using $ u = 1/r$, the above equation takes the form as
\begin{equation}
          \frac{d^2u}{d\phi^2}+u = \frac{M}{p^2} +
          3Mu^2 + \frac {aH}{2p^2} u^{(a-1)}+ \frac {(a+2)H}{2} u^{(a+1)}+ \frac {bQ}{2p^2} u^{(b-1)}
          + \frac  {(b+2)Q }{2} u^{(b+1)}
\end{equation}

where,

1)  For the case 3.1 i.e. corresponding solution (21), $H
=A/(3-\beta)$, $a = \beta-2$, $Q =B \beta/(4-\beta)(1-\beta)$, $b
= \beta-3$, $ B = $ proportional constant, $\beta = 2(
1+\frac{1}{\alpha})$, $A/8\pi =$ integration constant.

2)  For the case 3.2 i.e. corresponding solution (33), $H =8 \pi
D_0( 1+D)^2/(3+3D-\beta)$, $a =(3+3D-\beta)/(1+D)$, $Q = 0$, $ D =
$ proportional constant, $\beta = 2( 1+\frac{1}{\alpha})$, $D_0 =$
integration constant.

3)  For the case 3.3 i.e. corresponding solution (39), $H
=\Lambda/3$, $a = -2$, $Q =A/(3-\beta)$, $b = \beta -2 $, $
\Lambda = $ cosmological constant, $\beta = 2(
1+\frac{1}{\alpha})$, $D_0 =$ integration constant, $A=8 \pi D_0$.

Since the planetary orbits are nearly circular, so to analyze the
perihelion shift, we take a perturbation from the circular
solution, $u=u_0$, where

\begin{equation}
u_0 = \frac{M}{p^2} +
          3Mu_0^2 + \frac {aH}{2p^2} u_0^{(a-1)}+ \frac {(a+2)H}{2} u_0^{(a+1)}+ \frac {bQ}{2p^2} u_0^{(b-1)}
          + \frac  {(b+2)Q }{2} u_0^{(b+1)}.
\end{equation}

Now we assume, $u=(1+\epsilon)u_0$, where $\epsilon << 1$ and
substitute this in equation (46) retaining to the first order in
$\epsilon$, we get
\begin{eqnarray}
          \frac{d^2 \epsilon }{d\phi^2}= \left[6Mu_0 - 1 +
      \frac {a(a-1)H}{2p^2} u_0^{(a-2)}+ \frac {(a+2)(a+1)H}{2} u_0^{a} \right.\nonumber\\ + \left. \frac {b(b-1)Q}{2p^2} u_0^{(b-2)}
           + \frac{(b+2)(b+1)Q}{2} u_0^{b}\right]\epsilon.
\end{eqnarray}

From Einstein's general relativity, the precession coming from the
solar mass $M=M_\odot$ is

\begin{equation}
        \bigtriangleup \phi_0 = 6\pi M_{\odot}  u_0.
\end{equation}

In addition, there is a string density (with varying $\Lambda $)
induced precession as

\begin{eqnarray}
        \bigtriangleup \phi_{string} =  \frac {a(a-1) \pi H}{2p^2} u_0^{(a-2)}+ \frac {(a+2)(a+1) \pi H}{2} u_0^{a}+
         \frac {b(b-1) \pi Q}{2p^2} u_0^{(b-2)} \nonumber \\
          + \frac{(b+2)(b+1) \pi Q }{2} u_0^{b}.
\end{eqnarray}

Since observed perihelion shift of Mercury or other planets are
slightly vary with general relativistic prediction i.e.
$\bigtriangleup \phi_0 $, so this discrepancy would be overcome
due to presence of string density with varying $\Lambda$ in solar
system. In future study one can adjust the parameters for $100$
percent accuracy with the experiment result.

\section{Effective attractive correction to Newtonian force due
to cosmic string density admitting varying $\Lambda$}

The gravitational potential can be expressed within the Newtonian
limit as $\Phi = e^{\nu}/2$ and consequently the classical
gravitational acceleration ($g$) for our models are given by
\begin{equation}
        g = \frac{M}{r^2} + \frac{Ha}{2r^{(a+1)}} + \frac{bQ}{2r^{(b+1)}}
            \label{Eq1}
          \end{equation}
where, the parameters $a$, $b$, $H$, $Q$ are given as above for
the three sets of solutions (i.e. corresponding equations (21),
(33) and (39)).

From equation (51 ), it is readily understood that one can get an
effective attractive correction to Newtonian force due to presence
of cosmic string density admitting varying $\Lambda$. It seems
that our models give some clue how cosmic string with varying
$\Lambda$ would modify Newtonian as well as Non-Newtonian gravity
in various aspects. At the same time, our models would help to
solve missing mass problem (i.e. unseen dark matter in the
galaxies).

Again, taking Hubble radius (which is of the order of $10^{28}$
cm) as the value of $r$ it is easy to see from equation (23) that
the value of $\Lambda$ is of the order of $10^{-56} cm^{-2}$ which
agrees well with its present value. Also by putting $\beta= 3$ and
$B = -6$ in equation (16) and neglecting that the term involving
$1/r^2$ it is easy to see that the present value of the string
density $\rho_{s}$ is of the order of $10^{-57}$. Moreover, the
cosmological term and the string density are directly proportional
to each other. In the case of equation (27) also $\Lambda$ and
$\rho_{s}$ are directly proportional and the present value of the
former is one order of magnitude larger than the latter. This
means that depending on the signature of $\rho_{s}$ the Universe
will accelerate or decelerate.

\section{Conclusions}
Our three major results in the present investigations are as
follows:

{\it Firstly}, it is observed that variable cosmological constant
has power law dependence in the form $\Lambda=3r^{-2}$ for a
particular value for $\beta=3$. This gives the toroidal fluid
solution embedded in a locally anti-de Sitter
exterior~\cite{kri2003}.

{\it Secondly}, it is possible to show that by the suitable choice
of the values of the constants one can easily obtain the relation
$\Lambda=-8\pi{\rho}_s$. This means that locally vacuum energy
density and cosmic string density are interdependent to each
other. It is already known that cosmological constant is
responsible for providing repulsive pressure to the present state
of accelerating Universe~\cite{per1998,rie1998}. Therefore, one can explore the
possibilities of whether there is any relation between the cosmic
acceleration and cosmic string.

{\it Thirdly}, it is shown that cosmic string density can be
scaled by a power law of the type ${\rho}_s \propto
r^{-2}$~\cite{vil1984,tur1984,kib1986,got1987,dab1989}. This result
indicates that string-dominated universe expands more rapidly than
matter-dominated universe (${\rho}_m \propto r^{-3}$) or
radiation-dominated universe (${\rho}_r \propto r^{-4}$)~\cite{spe1997}.

It is also observed that from the general solutions several known
solutions such as the Schwarzschild vacuum solution, the
Schwarzschild-de Sitter vacuum solution, the Reissner-Nordstr{\"o}m
space-time and the Reissner-Nordstr{\"o}m-de Sitter universe can be
recovered here~\cite{sol1995}.

However, as concluding remarks we would like to give emphasis here
that only few of the above solutions seems to have some
relationship with nature, the rest are interesting only from the
mathematical point of view. It seems, for these few solutions with
some contact with the reality, a comparison with observations
would be a reasonable plan for validity of the features.
Therefore, we have proposed a test of our models to the Solar
System scales. For the same reason, it is possible to have an
estimate of the string density needed to have some interesting
deviations from the classical Newtonian potential and how these
values affect cosmological expansion.

\section*{Acknowledgments}
One of the authors (SR) would like to express his gratitude to the
authorities of IUCAA, Pune, India and ICTP, Trieste, Italy for
providing him Visiting Programmes under which a part of this work
was carried out. We are thankful to S. Randjbar-Daemi and P.
Creminelli, ICTP, Trieste, Italy for helpful discussions. Thanks
are also due to the referee for valuable suggestions which have
enabled us to improve the manuscript substantially.\\

{}

\end{document}